\documentclass[12pt, a4paper]{article}
\usepackage[utf8]{inputenc}
\usepackage{dcolumn,lscape}
\usepackage{amsmath,longtable,multicol,dcolumn,tabularx,graphicx,amssymb}
\usepackage{exscale,amsthm,multirow,rotating,subcaption}
\usepackage{natbib}
\usepackage[table]{xcolor}
\usepackage{bbm}
\usepackage{hyperref}
\usepackage{tikz,dsfont,float}
\definecolor{col1}{HTML}{EBEBDE} 
\definecolor{col2}{HTML}{777764} 
\definecolor{col3}{HTML}{4F4747} 

\usepackage{caption}
\usepackage{subcaption}
\usepackage{threeparttable}
\usepackage{booktabs}

\usetikzlibrary{patterns}

\usepackage{algorithm}
\usepackage{algpseudocode}
\begin{document}

\def\spacingset#1{\renewcommand{\baselinestretch}%
{#1}\small\normalsize} \spacingset{1}


\title{Spatio-temporal Event Studies for Air Quality Assessment under Cross-sectional Dependence}
\author{Paolo Maranzano\footnote{University of Milano-Bicocca (Milano, Italy) \& Fondazione Eni Enrico Mattei (Milano, Italy); paolo.maranzano@unimib.it}, Matteo Pelagatti\footnote{University of Milano-Bicocca (Milano, Italy); matteo.pelagatti@unimib.it} }

\maketitle

\begin{abstract}
Event Studies (ES) are statistical tools that assess whether a particular event of interest has caused changes in the level of one or more relevant time series.
We are interested in ES applied to multivariate time series characterized by high spatial (cross-sectional) and temporal dependence.
We pursue two goals.
First, we propose to extend the existing taxonomy on ES, mainly deriving from the financial field, by generalizing the underlying statistical concepts and then adapting them to the time series analysis of airborne pollutant concentrations.
Second, we address the spatial cross-sectional dependence by adopting a twofold adjustment. Initially, we use a linear mixed spatio-temporal regression model (HDGM) to estimate the relationship between the response variable and a set of exogenous factors, while accounting for the spatio-temporal dynamics of the observations. Later, we apply a set of sixteen ES test statistics, both parametric and nonparametric, some of which directly adjusted for cross-sectional dependence.
We apply ES to evaluate the impact on NO$_2$ concentrations generated by the lockdown restrictions adopted in the Lombardy region (Italy) during the COVID-19 pandemic in 2020. The HDGM model distinctly reveals the level shift caused by the event of interest, while reducing the volatility and isolating the spatial dependence of the data. Moreover, all the test statistics unanimously suggest that the lockdown restrictions generated significant reductions in the average NO$_2$ concentrations.
\end{abstract}

\noindent%

\spacingset{1.45} 

\newpage

\section{Introduction} \label{Sec:Introduction}

\subsection{Event Studies and Statistical Intervention}
Event studies, hereafter ES, are statistical tools used to assess whether a particular event of interest induced abrupt changes in the level (level shift) or in the volatility (volatility shift) of one or more relevant time series \citep{CampbellLoMacKinlay1998}. Such events may be either artificial, such as the introduction of a new policies or interventions (e.g., the effect of traffic limitations on road traffic and air quality or restrictive measures on mobility), or natural events (e.g., environmental extreme events).

ES ground on two main pillars. The first pillar is the \textit{interrupted time series} (ITS) paradigm \citep{McDowallEtAl2019}, in which we assume that at a certain known date an event (e.g., treatment or intervention) occurred, dividing the time series into two parts, one before and one after the event. The occurrence date is labeled as the \textit{event date} and is assumed to be known. The final goal of an ES is to state if the event of interest generated a statistically significant impact on the time series under consideration. The event can be permanent or temporary and it ca be gradual or abrupt. Usually, ES focus on abrupt changes.
The second pillar is the \textit{offline hypothesis testing} methods of \cite{Basseville1993}, in which a \textit{without change} scenario (i.e., no abrupt changes occurred in the observed data) is compared to a \textit{with change} scenario (i.e., the data were statistically significantly affected by some shock or event). The two scenarios are compared through a statistical hypothesis testing procedure. Considering ES for level shift, under the null hypothesis H$_0$, we state that there are no abnormal variations in the average level due to the event occurrence, whereas the alternative H$_1$ suggests the existence of abnormal variations after the event of interest. In the former case, the variations can be attributed to the randomness in the data.

ES are directly connected with branches of statistics involved into the impact assessment of policies and unexpected shocks, such as the Statistical Intervention Analysis \citep{Abraham1980}, hereafter SIA. Indeed, ES and SIA can be considered as complementary, but also competing, tools.
ES methodology combines a regression-based approach for parameter estimation enforced with a validation strategy and robust hypothesis testing to return insights about the sign of the shock and its statistical significance. SIA is used to quantify the impact of the event through a model-based inferential approach. However, the average effect generated by the event is estimated by regression techniques and its statistical significance is assessed by classic hypothesis tests (e.g., t-test for regression coefficients).
In ES, the event impact is quantified as pointwise change (i.e., at a given time instant) or cumulative change (i.e., over a window of subsequent observations) in the time series of interest.
Both ES and SIA are based on the ITS paradigm, however, whereas in ES much emphasis is placed on robust estimation of the significance of variances, statistical intervention places great emphasis on defining the form of the intervention and on ensuring an unbiased estimate of the effect.
Accordingly, it is reasonable to consider the combined use of ES and SIA for policy impact assessment. For example, one can adopt ES as a preliminary tool to identify the presence (or absence) of shocks in the time series and then quantify the change that has occurred using statistical intervention tools.

Both ES and SIA follow a model-based reasoning, but adopt very different strategies for computing the test statistics. 
ESs for level shifts follow a forecasting-based approach in which the level before and after an intervention are compared to assess whether the estimated difference is statistically significant. The ES test statistics are computed using the forecast errors on future observations at the event date. Such approaches are standard, for example, in assessing structural breaks in public and macroeconomic policies \citep{GuglielminettiRondinelli2021}.
The standard procedure for ES is to segment the timeline into two consecutive subsamples: the first part of the time series is used to estimate the parameters of the regression model (i.e., \textit{the estimation window}), while the second part is used to quantify the effect of the intervention without re-estimating the model (i.e., \textit{the event window}). To produce unbiased parameter estimates, the estimation window must be unaffected by the intervention/event.
For level shift ESs, if the event truly generated a significant effect, post-event observations should diverge from model-based forecasts. Therefore, the forecast (residual) errors should exhibit a pronounced change in the level shift. In contrast, if the variation is random, the forecast and actual values should almost coincide and the residuals should have zero mean. 
From the sampling perspective, ES approach is similar to modern statistical learning techniques for temporal data, in which a portion of the time series is used for in-sample model training and parameter estimation, whereas performance is evaluated over successive out-of-sample periods in cross-validation \citep{Hyndman2018}.
While ES use a validation strategy, in SIA the model's parameters are estimated employing all available observations, that is, the entire time series. The classic approach to SIA identifies the level (or volatility) shift in time series using a regression framework in which an intervention covariate is introduced to quantify the anomaly. The covariate can take the form of a dichotomous variable (pulse or constant effect), ramp (linear effect), or dynamic ARMA-like filter \citep{BoxTiao1975,Abraham1980}. If a model in parametric form is fitted, the significance of the event is reflected by the statistical significance of the parameter associated with the intervention covariate.


\subsection{Event Studies for Environment and Energy} \label{subsec:ES_literature}
We focus on the analysis of air quality data and the application of ES techniques to detect shocks in observed airborne pollutant concentrations following unexpected exogenous events.
While statistical intervention analyses are common tools in studying the air quality \citep{Fasso2013,Grange2019,Petetin2020} and the impact of pollution mitigation policies, ES are only recently receiving attention in pollution-related fields, such as energy and oil commodity markets. For example, \cite{DemirerEtAl2010} use the ES methodology to examine the behavior of crude oil spot and futures markets around the OPEC conference, as well as US strategic petroleum reserves announcement between 1983 and 2008. \cite{ZhangEtAl2009} use ES to test the impact of extreme events, such as the Gulf War in 1991 and  Iraq War in 2003, on crude oil price volatility. Further, in \cite{ZhaEtAl2018} the authors aim at assessing the impact of refined oil price adjustments to control air pollution in China between 2014–2015.
In addition, ES methods have recently received great attention in climate policy analysis. Looking at the macroeconomic perspective, one can refer to the paper by \cite{Barnett2019}, in which the impacts of climate policy risk exposure on observable market outcomes such as oil production, stock returns, and oil prices are analyzed. In \cite{DiazRaineyEtAl2021}, the authors examined the effect of policy interventions associated with Paris Agreement (agreement and ratification) and the election of Donald Trump (election and withdrawal from agreement) on stock returns of oil and gas companies. Other researchers focused on the effect of climate polices on stock returns and investment portfolios. We recall, for example, \cite{BorghesiEtAl2022} examining the behavior of green and brown portfolios around green policy-related announcements launched by European governments in 2020 to alleviate the adverse effect of climate change; \cite{BirindelliChiappini2021}, which examined investor reaction toward eight EU policy announcements over the years 2013–2018 on a large sample of EU firms; and \cite{HuynhXia2020}, which used ES to analyze the effect of climate change news on individual corporate bond returns.

None of the previously cited studies directly examine the case of local pollutant concentrations. Moreovor, they employ standard regression techniques for time series under the assumption of cross-sectional independence (i.e., the series are independent of each other). In this paper we are interested in ES applied to multivariate time series characterized by high cross-sectional (spatial) and temporal dependence. In particular, we are interested in concentrations of airborne pollutants observed on geo-referenced monitoring networks located on specific territories. In this case, cross-sectional dependence (CD) is a direct consequence of the spatial correlation between the sampling points in space. Indeed, it is reliable to assume that control units located at close distances record similar values under the same environmental conditions.

Since the data are characterized by a well-defined spatial and temporal structure, a key feature of this paper is the implementation of an appropriate geostatistical modeling approach to model the relationships between the pollutant concentrations across space and time. In addition, the focus is on the statistical significance of observed variations in the average value under autocorrelated data across space and time, and on the methods used to correct for these issues. Quantification of impact is a secondary objective.

Specifically, we aim at targeting two issues.
The first concerns the definition of Event Studies specific to the case of air quality.
In this regard, we propose to extend the existing taxonomy on ES, mainly deriving from the financial field, by generalizing the underlying statistical concepts and then adapting them to the case of airborne pollutant concentrations. 
This task is performed by assimilating the concentrations of a specific pollutant observed in a given time $t$ at a given monitoring station $s$ to the stock returns observed for a stock $i$ at time $t$. Thus, by analogy with the definitions of normal and abnormal returns \citep{BallBrown1968}, we propose the definition of normal, abnormal, and cumulated abnormal airborne pollutant concentrations.
The second issue addresses the problem of adjusting ES against cross-correlation when dealing with spatially distributed observations. To address the spatial cross-sectional dependence, we adopt a twofold adjustment with respect to classical ES frameworks. First, we use a linear mixed spatio-temporal regression model, namely the Hidden Dynamics Geostatistical Model or HDMG \citep{Calculli2015}, to estimate the relationship between observed concentrations and several exogenous factors, such as meteorology and calendar effects, while accounting for the spatio-temporal dynamics of the data. To exploit the improvements guaranteed by the spatio-temporal modelling, the HDGM is compared with purely temporal regression models (i.e., not adjusted for spatial dependence).
Second, we apply and compare a series of sixteen ES test statistics, both parametric and nonparametric, some of which directly adjusted for cross-sectional dependence. Expectations are for greater consistency in corrected statistics than in unadjusted ones.
Extending robust ES to the spatio-temporal context is an improvement that can provide great benefits to projects whose goal is to assess the impact of policies in space and time. A valid example is the AgrImOnIA project (\href{https://agrimonia.net/}{https://agrimonia.net/}), whose objectives include the analysis of the relationship between livestock and air pollution through geostatistical models, as well as scenario analyses regarding the expected effect on particulate matter concentrations deriving by reductions of pollutant emissions in the agricultural sector (e.g., efficient manure management, increased organic farming techniques).

The remainder of this paper is organized as follows. In Section \ref{sec:ES_AQ}, we propose the definitions of normal (expected) concentrations, abnormal (unexpected) concentrations, and cumulated airborne pollutant concentrations.
In Section \ref{sec:ES_crosscorr}, we discuss the effect of CD on the test statistics and present the twofold adjustment strategy for CD data in ES. In Section \ref{sec:HDGM}, we present the HDGM and discuss its interpretation and the major benefits of its implementation.
In Section \ref{sec:ES_application}, we present an empirical application of ES analysis concerning the effect of the lockdown restrictions imposed on air quality in the Lombardy region (Italy) in response of COVID-19 disease spread in 2020. The aim was to test whether traffic and mobility restrictions significantly affected the average daily levels of NO$_2$ concentrations collected at 83 ground monitoring stations. The results support the hypothesis of a generalized reduction of the average oxide concentrations in the region.
Finally, Section \ref{sec:ES_conclusions} concludes the paper and proposes future development strategies for ES in spatio-temporal frameworks.

\section{Event Studies taxonomy for air quality assessment} \label{sec:ES_AQ}
Let $s=1,\ldots,N$ be the index for cross-sectional units, where $N$ is the total number of spatial locations in the sample, and let $t$ be the time index $t=T_0,\ldots,T_1,T_1+1,\ldots,T$.
Recall that ES rely on a validation strategy performed by splitting the whole temporal sequence into two subsamples, namely the estimation window and the event window. Observations in the estimation window are used to estimate the model parameters, while those in the event window are used to assess the event's effect.

Let's denote the time index for the first available observation as $T_0$, the time index for the last observation included in the estimation window as $T_1$, and the time index for the last observation in the event window with $T$. Moreover, assumes that the event date coincides with $T_1+1$ and that the last observation of the event window coincides with the last available observation $T$. The set $\Omega_0 = T_0 + 1, T_0 + 2,\ldots, T_1$ (with cardinality $\tau_0$) will denote the set of time indexes for the estimation window, while the set $\Omega_1 = T_1 + 1, T_1 + 2,\ldots, T$ (with cardinality $\tau_1$) will denote event window. Eventually, the set $\Omega = \Omega_0 \cup \Omega_1$ has cardinality $\tau = \tau_0 + \tau_1$ and represents the full sequence of time indexes.

Let C$_{st}$ be the observed airborne pollutant concentrations observed at time $t$ at the monitoring station $s$. Moreover, let X$_{st}$ be a set of conditioning information collected at time $t$ at the same station $s$. Notice that the set X$_{st}$ can include station specific information (e.g., local weather measurements, traffic data, land cover) and information common to all the sensor (e.g., calendar effects) or further variables measured at aggregate geographical level. Assuming the existence of a statistical relationship between the concentrations and conditioning information, for each spatial sampling point $s$ and time stamp $t$, the normal concentrations (NC$_{st}$) can be defined as the conditional expected value of C$_{st}$ with respect to X$_{st}$, i.e.
\begin{equation}
NC_{st} = E[C_{st} \rvert X_{st}] \qquad \forall t \in \Omega.
\end{equation}
It follows that the abnormal concentrations at time $t$ for station $s$ (AC$_{st}$ or $\varepsilon_{st}$) are defined as the difference between the observed concentrations at time $t$ and the normal concentrations at time $t$ for station $s$:
\begin{equation}
AC_{st} = \varepsilon_{st} = C_{st} - E[C_{st} \rvert X_{st}] = C_{st} - NC_{st} \qquad \forall t \in \Omega.
\end{equation}
The abnormal concentrations in the event window $\Omega_1$ can be interpreted as the abnormal values of $C$ not explained by the conditioning information X$_{st}$, and potentially generated by the event of interest.


Finally, by the term cumulated abnormal concentrations we mean the cumulative sum of abnormal concentrations in a given time window. When the cumulative window coincides with the event-window, we refer to cumulated abnormal concentrations in the event window, hereafter CAC$_{s\Omega_1}$.  The cumulated abnormal concentrations for station $s$ over the event window $\Omega_1$ are defined as follows:
\begin{equation} \label{eq_est_cac}
CAC_{s\Omega_1} = \sum_{t = T_1 +1 }^{T_2}{AC_{st}}.
\end{equation}

\section{Addressing cross-sectional dependence in ES} \label{sec:ES_crosscorr}
Cross-sectional dependence is the outcome of the strong spatial correlation naturally exhibited by observations collected on geo-referenced monitoring networks located on specific territories. Recall the Tobler's First Law of Geography \citep{Tobler1970}, \textit{'Everything is related to everything else, but near things are more related than distant things'} (spatial autocorrelation principle). The law also applies to air quality, as it is reasonable to assume a strong spatial correlation in air quality of geographically close areas \citep{MonteroEtAl2021}. Also, recall the Third Law of Geography \citep{ZhuTurner2022} \textit{‘The more similar geographic configurations of two points, the more similar the values (processes) of the target variable at these two points’}.
Therefore, by analyzing the measurements recorded in monitoring networks belonging to specific regions and assuming very similar environmental conditions, it is realistic to state that the concentrations of airborne pollutants observed in monitoring stations located at close distances will be very similar.

According to \cite{DaleFortin2009}, ecological data are often affected by positive spatial dependence, giving rise to more apparently significant results than the data justify. When the observations are affected by positive cross-sectional dependence, the significance rates of the classic ES parametric test statistics are oversized (i.e., the test rejects too often). On the contrary, in the presence of negative cross-sectional dependence ES tests are undersized (i.e., the test rejects too few). The same considerations hold when observations are affected by temporal autocorrelation \citep{LeeLund2004,LeeLund2008}, or when considering correlated paired samples \citep{DutilleulEtAl1993,Zimmerman2012}. Size-distortion effects still hold when negligible levels of cross-correlation are observed \citep{PelagattiMaranzano2021}.
Considering ES applied to environmental policy evaluation, over-rejection leads to the erroneous identification of a statistically significant effect caused by the event of interest. Certainly, using regression techniques to compute the abnormal concentrations can reduce the average correlation significantly. However, because of the strongly similar environmental and geographical conditions \citep{ZhuTurner2022} of the region, even after filtering the confounding effects there may be enough average correlation to severely bias any parametric statistic.

To address the spatial cross-correlation issue, we adopt a twofold adjustment compared to classic ES frameworks.

The first adjustment concerns the modelling (regression) step.
Usually, ES literature assumes a linear relationship among the response variable and the covariates. Moreover, the relationship is often estimated in a univariate framework, that is, for each location $s$ the observed concentrations are modelled as linear functions of the set of conditioning information X$_{st}$. This is equivalent to assuming that the concentrations observed at the stations are mutually independent (which is rather unlikely given the physical-chemical characteristics of the atmospheric phenomenon). Also, in linear regression contexts, estimates of regression coefficients could be biased due to not explicitly modelled spatial structures into the residuals \citep{Paciorek2010}. Since ES statistics are calculated from the predicted residuals, the presence of confounding spatial correlation can adversely affect the values of the statistics and their uncertainty.
We aim at relaxing the independence assumption by explicitly modelling the spatio-temporal dynamics of the NO$_2$ concentrations though a common spatio-temporal process describing the evolution of airborne pollutant concentrations measured on the network. We use a linear mixed spatio-temporal regression model, namely the Hidden Dynamics Geostatistical Model (HDMG) \citep{Calculli2015}, to estimate the relationship between observed concentrations and several exogenous factors, such as meteorology and calendar effects, while accounting for the spatio-temporal dynamics between the observations.

The second adjustment refers to the hypothesis testing step.
In line with the reasoning of \cite{LeeLund2008}, in \cite{DaleFortin2002} and \cite{DaleFortin2009} the authors proposed a strategy to mitigate the CD effect by adjusting the test statistics for the effective sample size, that is the extent to which the test statistic must be deflated to achieve the correct rejection rate. The strategy we adopt is related to the one proposed in \cite{Zimmerman2012} for t-tests on two related populations. Rather than operating on the information contained in the observations (i.e., effective sample size), we propose to use CD-adjusted test statistics directly adjusting for spatial cross-sectional dependence by means of a cross-correlation measure for multivariate time series.
We apply and compare the set of ES test statistics presented and discussed in \cite{PelagattiMaranzano2021}. The proposed statistics test the null hypothesis of absence of level shift in the cumulated abnormal concentrations ($CAC$). The list of ES we used includes both cross-sectionally adjusted statistics and non-adjusted statistics. Moreover, among the considered statistics, we included both parametric and non-parametric specifications. In particular, we considered nonparametric statistics belonging to the family of rank-based statistics \citep{KolPyn2011,Luoma2011,HagnasPynnonen2014}.
Since we model the spatio-temporal dependence at the estimation stage, the residual cross-correlation is expected to be very small. Therefore, all the adjusted statistics should be straightforward in identifying (or not identifying) the statistical significance of the effect. Statistics that do not account for the dependence should nevertheless be robust, but not as robust as the adjusted statistics.

In Table \ref{tab:ES_AQ_stats} we list the ES test statistics proposed or analyzed by \cite{PelagattiMaranzano2021} and that we use in the empirical analysis. The table includes the name of the statistics, a bibliographical reference and a statement regarding the adjustment for cross-sectional dependence. An extended discussion on the statistical properties, as well as simulated and empirical results about their performances, of each statistic is available in the same article.

\begin{table}[htbp]
\centering
  \begin{threeparttable}
  \caption{Test statistics for H$_0$: $E[CAC_{\Omega_1}] = 0$}
  \label{tab:ES_AQ_stats}%
  \footnotesize
\begin{tabular}{cccc}
	\toprule
	   Cross-sectional & \multirow{2}{*}{Specification} & \multirow{2}{*}{Statistic} & \multirow{2}{*}{Reference} \\
	   adjustment       &               &           &           \\
	\midrule
	Not-adjusted & Param. & cross\_T\_Test & \cite{BroWar1985} \\ 
	Not-adjusted & Param. & crude\_dep\_T\_Test & \cite{BroWar1985} \\ 
	Not-adjusted & Param. & T\_skew & \cite{BroWar1985} \\
	Not-adjusted & Param. & Z\_patell & \cite{Pat1976}\\ 
	Not-adjusted & Param. & Z\_BMP & \cite{Boeetal1991}\\ 
	Not-adjusted & Nonpar. & Z\_grank & \cite{KolPyn2011}\\ 
	Not-adjusted & Nonpar. & CumRank\_Z & \cite{Cor1989} \\
	Adjusted & Param. & Z\_patell\_adj & \cite{KolPyn2011} \\  
	Adjusted & Param. & Z\_BMP\_adj & \cite{KolPyn2011} \\
	Adjusted & Nonpar. & T\_grank & \cite{KolPyn2011}\\  
	Adjusted & Nonpar. & Z\_grank\_adj & \cite{KolPyn2011}\\
	Adjusted & Nonpar. & CumRank & \cite{Cor1989}\\  
	Adjusted & Nonpar. & CumRank\_mod & \cite{CorZiv1992}\\  
	Adjusted & Nonpar. & CumRank\_T & \cite{CorZiv1992} \\ 
	Adjusted & Nonpar. & CumRank\_Z\_adj & \cite{HagnasPynnonen2014} \\  
	Adjusted & Nonpar. & P$_1$ & \cite{PelagattiMaranzano2021} \\  
	Adjusted & Nonpar. & P$_2$ & \cite{PelagattiMaranzano2021} \\  
	Adjusted & Nonpar. & Corrado-Tukey adjusted & \cite{PelagattiMaranzano2021} \\  
	\bottomrule
\end{tabular}
  \end{threeparttable}
\end{table}

Spatial dependence can be measured in several ways (we suggest referring to the review paper by \cite{Lee2017} for a detailed discussion about strengths and weaknesses of autocorrelation and bivariate spatial correlation measures). The most intuitive measure of correlation over space is the Moran's Index \citep{Moran1950}, which can be seen as the correlation coefficient for the relationship between a variable and its neighbouring values. Global Moran's I is a generalization of Pearson's correlation coefficient with geographical weights \citep{Chen2013}.
The very close analytical link between Pearson's coefficient and Moran's index was first scrutinized by \cite{Wartenberg1985}. Later, a bivariate spatial correlation measure was proposed in \cite{Lee2001} as a smoothed version of Pearson's correlation capable of capturing both bivariate \textit{point-to-potint association} and univariate spatial autocorrelation.
The CD adjusted statistics in Table \ref{tab:ES_AQ_stats} correct the spatial dependence using Pearson's linear correlation coefficient.
While this is an approximate \textit{aspatial} measure of autocorrelation, it can still provide a straightforward indication of the direction (sign) of the relationship and its strength. Indeed, the bivariate-spatial correlation can be expressed as a fraction of Pearson's linear correlation coefficient, which acts as an upper-bound \citep{Lee2001}.

\section{HDGM for Event Studies} \label{sec:HDGM}
We assume that data are generated by a spatio-temporal process $\{Y_{st} \in \mathds{R} : s \in D, t = 1, \ldots, T \}$, where $D$ is the spatial domain and $t$ represents a discrete point of time. The Hidden Dynamics Geostatistical Model, namely the HDGM \citep{Calculli2015}, entails a random-effect term $w_{st}$ modelling the spatial and temporal dependence, a fixed-effect term $v_{st}$ accounting for all exogenous regressive effects and an error term $\varepsilon_{st}$. The univariate HDGM is specified as follows:
\begin{equation}\label{eq:HDGMmodel}
  Y_{st} = v_{st} + w_{st} + \varepsilon_{st}
\end{equation}
with $\varepsilon_{st} \sim N(0,\sigma^2_{\varepsilon})$ being the error vector that is assumed to be independent and identically distributed (I.ID.) across space and time. The error sequence is supposed to have a zero mean and a constant variance $\sigma^2_\varepsilon$ and Gaussian distribution. The fixed-effects mean term can be specified as follows:
\begin{equation}\label{eq:HDGMfixedeffects}
 v_{st} = X_{st}\beta ,
\end{equation}
where $X_{st}$ is the $(n \times p)$-dimensional matrix of $p$ covariates at location $s$ and time point $t$ and $\beta$ is a vector of $p$ fixed-effects parameters.
The random effects term $w_{st}$, which accounts for the spatio-temporal dependence in the random process $Y_{st}$, is described by a Markovian autoregressive temporal process plus a spatially correlated random effects $\omega_{st}$ with Gaussian distribution
\begin{equation}\label{eq:HDGMrandomeffects}
 w_{st} = g w_{st-1} + \omega_{st} \,
\end{equation}
where $\lvert g \rvert < 1$ represents the first-order temporal autocorrelation parameter. The element $\omega_{st} \sim N(0,\Gamma)$ is a Gaussian Process independent in time with zero mean and covariance matrix
\begin{equation}\label{eq:HDGMrandomeffects2}
 \Gamma(s,s') = \nu \rho(\lvert\lvert s - s' \rvert\rvert, \theta) \,
\end{equation}
where $\nu > 0$ is the random effects variance (shape parameter), $\rho$ is a Matérn spatial covariance function which depends on the distance between pairs of observations, i.e. $\lvert\lvert s - s' \rvert\rvert$, and $\theta > 0$ is the spatial dependence parameter.
HDGM allows both univariate and multivariate specifications \citep{SPASTA2021}, but also a functional specification called f-HDGM \citep{DSTEM22021}. Each specification relies on a state-space representation with the related Kalman Filter algorithm. For recent reviews on Kalman filtering for spatio-temporal models see \cite{FerreiraMateuPorcu2022}, \cite{JurekKatzfuss2022}, and \cite{JurekKatzfuss2022_Envir}.
The maximum likelihood estimates of the parameters are computed using the EM algorithm, which is implemented together with the parameter variance-covariance matrix computation in D-STEM package \cite{DSTEM22021} for MATLAB.

The HDGM has been extensively used in many empirical contexts, such as air quality policy assessment \citep{SPASTA2021}, bike-sharing system comprehension \citep{PiterOttoAlkhatib2022}, and off-shore coastal profile measurements for beach monitoring \citep{OttoPiterGijsman2021}. In addition, the HDGM model, combined with land use regression techniques, has been used in \cite{TaghaviEtAl2019} for spatio-temporal interpolation of missing observations outperforming conventional interpolation techniques.

\section{Assessing the impact of COVID-19 lockdown measures on air quality in Lombardy} \label{sec:ES_application}
To fight the spreading COVID-19 disease across the country, the Italian government imposed a total lockdown \citep{DPCM8marzo} from 9$th$ March to 18$th$ May 2020. This period, also denoted as \textit{first wave COVID-19 lockdown}, was characterized by the closure of all non-essential activities and enterprises, and by the minimization of individual mobility and social distancing \citep{HP2021}. As a direct consequence of the limitations, a generalized reduction of car traffic and personal travel for the entire country was observed \citep{FinazziFasso2020}.

Numerous studies have shown how general lockdown imposed by governments have generated strong and significant reductions in pollutant concentrations worldwide \cite{Higham2020,Zangari2020,Nakada2020,XinYalu2021}, particularly in large urban centers \citep{Baldasano2020,Rossi2020}. The Lombardy (Northern Italy) case study received a significant scientific interest. In particular, the studies by \cite{Collivignarelli2020,Cameletti2020,SPASTA2021,MaranzanoFasso2020,Granella2021} showed that, due to the restrictions on mobility, oxides concentrations registered statistically significant reductions (up to 50\%) throughout the region. On the contrary, particulate matter remained stable or slightly reduced over the entire period. This indicates that the major emission sources of particulate matter in the region are other than vehicular traffic and industrial production. Consider, for example, the role of agriculture and livestock farming, which, through the production of ammonia, generates significant amounts of secondary particulate matter \citep{Lovarelli2020,Lovarelli2021}.


\subsection{Sample and event definition}
We are interested in analyzing the effect of the lockdown restriction on NO$_2$ concentrations registered in Lombardy. 
The null hypothesis we are testing is that the restrictions did not have any effect on the cumulative abnormal NO$_2$ concentrations during the lockdown period (i.e., $CAC_{s\Omega_1}$ have null mean value). The alternative hypothesis is that the cumulated abnormal concentrations registered a significant reduction during the event-window (i.e., $CAC_{i\Omega_1}$ have negative mean value). Therefore, the hypothesis test is unilateral on the left tail.
Our expectations about the tests are twofold. Previously existing literature confirmed significant reductions in NO$_2$ levels in Lombardy. Thus, a statistically significant negative sign of the statistics is expected. As for dependence, we expect the cross-sectionally adjusted statistics to be moderate in absolute terms when compared to the unadjusted statistics. As we will see later, the empirical correlation of pollutants is remarkable, so that the unadjusted statistics tend to lead to over-sized tests.

We considered the average daily concentrations of NO$_2$ collected in $N = 83$ ground stations belonging to the ARPA Lombardia network \citep{Maranzano2022}. Air quality measurements were collected using the \textit{ARPALData} package (release 1.2.2 available on CRAN) of the statistical software \textit{R} \citep{RCore2020}.

We performed ES analysis using daily observations from 1st January 2018 to 18th May 2020. We recall that the lockdown period during the first wave of COVID-19 lasted 70 days, from 9$th$ March and 18$th$ May 2020. Although the event window is very long, we can credibly assume that there were no other overlapping events capable of hiding the impact of the restrictive measures during the period. Moreover, by controlling through regression for possible exogenous effects, several confounding factors typical of environmental series, such as changes in meteorology or calendar events, were isolated. However, as a further test regarding the choice of the windows, we performed a sensitivity analysis by estimating the models with various combinations of estimation window and event window. We evaluated the following scenarios:
\begin{itemize}
    \item Exact lockdown period: estimation window from 1st January 2018 to 8th March 2020 ($\tau_0 = 798$ days) and event window from 9th March 2020 to 18th May 2020 ($\tau_1 = 70$ days)
    \item Anticipated lockdown period: estimation window from 1st January 2018 to 29th February 2020 ($\tau_0 = 790$ day) and event window from 1st March 2020 to 18th May 2020 ($\tau_1 = 78$ days)
    \item Restricted lockdown period: estimation window from 1st January 2018 to 15th March 2020 ($\tau_0 = 805$ days) and event window from 16th March 2020 to 16th May 2020 ($\tau_1 = 61$ days)
\end{itemize}

The results of the three scenarios were mutually consistent. The diagnostics on the estimated models showed very similar characteristics and did not reveal structural differences in the time series. Both magnitudes and significance of the test statistics remained of the same order and were absolutely comparable. The overall assessment of lockdown effects remained unchanged.
Therefore, only the results referring to the exact lockdown period scenario, i.e., considering an estimation window between 1st January 2018 and 8th March 2020 and event window from 9th March 2020 to 18th May 2020, will be shown and commented on in the following sections. The full results of the three scenarios are available in Supplementary Material S1.

Table \ref{tab:descriptNO2} reports key descriptive statistics for the Pearson's linear correlation index ($\rho$) of NO$_2$ concentrations observed in the sample. The average cross-sectional correlation at regional level for NO$_2$ is about 60\%, thus able to heavily bias classical ES statistics \citep[see the discussion in][regarding the effect of even a small correlation on size and power of ES tests]{PelagattiMaranzano2021}.
\begin{table}[htbp]
 \centering
 \caption{Descriptive statistics for NO$_2$ concentrations observed at the 83 stations included in the sample and regarding the period January 2018 - May 2020.}
   \begin{tabular}{lrrrrrr}
   \toprule
                & Min  & 25\% perc & mean & Median & 75\% perc & Max \\
   \midrule
   $\rho$  & -0.327 & 0.464 & 0.579 & 0.632 & 0.737  & 0.953 \\
   \bottomrule
   \end{tabular}
 \label{tab:descriptNO2}
\end{table}

\subsection{Spatio-temporal modelling and diagnostics}
The covariates included in the fixed-effect component of the HDGM were chosen among those available from the Copernicus ERA-5 reanalysis database \citep{ERA5Land}. ERA-5 provides observations with a $0.1^\circ\times 0.1^\circ$ grid spatial resolution. To each air quality station, we associated the meteorological measurement observed in the cell where the station is located. To explain the airborne pollutant concentrations, we considered a set of nine meteorological and land cover variables: average daily temperature ($\mbox{}^{\circ} C$), daily cumulative precipitation (mm), relative humidity ($\%$), atmospheric pressure (Pa), eastward and northward component of the wind (m/s), geopotential height (m$^2$/s$^2$) as a proxy of altitude, and high and low vegetation covering  \citep[measured as one-half of the total green leaf area per unit horizontal ground surface area, cf.][]{ERA5Land}. While the geopotential height and land cover are time invariant, the weather covariates are all time-varying.

Pollutant concentrations are strongly seasonal phenomena. In particular, they consistently follow the cyclical pattern of climatic seasons. Statistically, they are characterized by annual seasonality and intra-weekly seasonality. In the case of hourly measurements, concentrations are also affected by intra-day seasonality. The inclusion of the weekend dummy in the list of covariates allows us to control for typical reductions observed during the weekend, while the use of temperature serves as a proxy for the climatic season. In the case of NO$_2$, the use of temperature is not sufficient to suitably capture the annual cycle. To resolve this issue, we decided to include short- and long-term lags of the time-varying covariates. In detail, we included the 1-day, 2-day and 365-day lagged values of temperature, rainfall, pressure, wind speed and relative humidity among the regressors. Altogether, the total number of covariates is 28.

In order to assess the effectiveness of spatio-temporal modelling, we compared the HDGM performances with those that would be obtained by ignoring the spatial dependence in the estimation stage, while keeping active the the temporal dynamics of the concentrations.
We compared the spatio-temporal model with two classes of univariate linear models. First, we considered a multiple linear regression with i.i.d.\ errors. Second we model the air quality through a linear regression model with ARMA disturbances, or regARMA \citep{ZindeGalbraith1991,Veleva2020,Vadrevu2020}.
Specifically, the regression models we considered are as follows:
\begin{itemize}
    \item \textbf{HDGM}: HDGM as defined in Equations (\ref{eq:HDGMmodel})-(\ref{eq:HDGMrandomeffects2})
    \item \textbf{regAR1}: regARMA with AR(1) dynamics
    \item \textbf{regARMA}: regARMA with optimal ARMA dynamics
    \item \textbf{lm}: linear regression with i.i.d.\ errors
\end{itemize}

Note that, while for the regAR1 model the ARMA structure is pre-determined, regarding the optimal regARMA the ARMA specification is automatically detected using the Corrected Akaike Information Criterion (AICc) at each monitoring site. Dealing with daily data we restricted the ARMA specifications to orders in the range ARMA(0,0)--ARMA(7,7). Thus, we allow the concentrations not explained by the regressors to be affected by their past values up to seven days and random shocks occurring within one week of the date.

Both the regAR1 and the linear regression model can be seen as special cases of the HDGM model. In fact, by assuming that the temporal dynamics of the errors is of type AR(1), and assuming absence of spatial correlation, HDGM becomes a regARMA(1,0) model with common (pooled) coefficients across stations. Finally, assuming both spatial and temporal incorrelation, the HDGM model degenerates into a pooled regression model with i.i.d.\ errors. Thus, lm, regAR1 and HDGM can be considered nested models. The optimal regARMA model is here included as we assume that the temporal autocorrelation is very strong and the use of a simple AR(1) dynamics can be insufficient to suitably capture the persistence.

Eventually, in order to maintain as much comparability as possible, we decided to use a common set of spatio-temporal covariates to be included in the fixed-effects component of each model.

\begin{figure}[htbp]
\includegraphics[width=\textwidth]{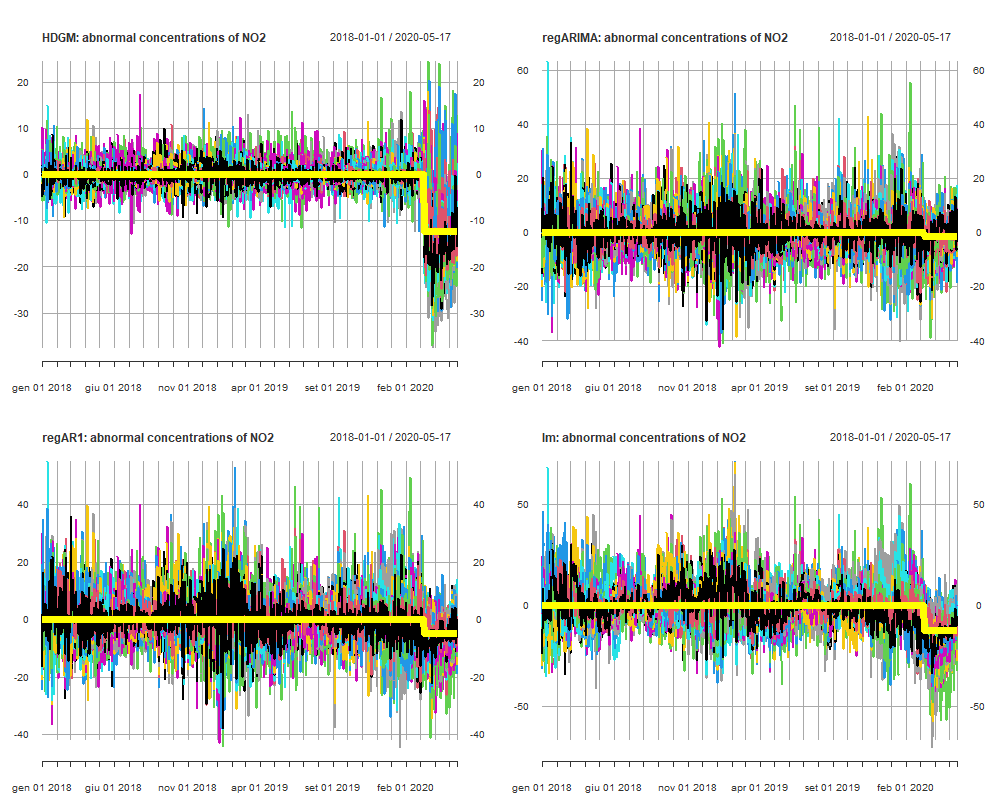}
\caption{Estimated abnormal concentrations using HDGM (upper left panel), optimal regARMA (upper right panel), regAR1 (lower left panel) and linear model (lower right panel). Estimation window: 1st January 2018 - 8th March 2022. Event window: 9th March 2022 - 18th May 2022.}
\label{fig:ES_stats}
\end{figure}

Figure \ref{fig:ES_stats} shows the estimated abnormal concentrations using the four models both in the estimation and event windows.
The insights provided by the chart are manifold. First, the HDGM model is the only one to distinctly reveal the level shift caused by the event of interest. There is a reduction of about -12$\mu g/m^3$ between the average of the estimation window and the event window. This difference is close to zero in the case of the optimal regARMA model and is estimated to be around -5$\mu g/m^3$ in the other two cases. The reduction estimated with HDGM is the only one consistent with that estimated in other studies \citep{LonatiRiva2021,BontempiEtAl2022}. Second, the abnormal concentrations of HDGM were the least volatile among the four. In fact, the range in the estimation window is between $\pm10 \mu g/m^3$ and is constant for the entire period (no heteroskedasticity). Moreover, the abnormal concentrations are poorly seasonal. Pure time series models show seasonality in the residuals, which are highly volatile. Graphical analysis leads unquestionably to the dominance of the spatio-temporal model over those that do not model spatial correlation.

Table \ref{tab:descript_AC_NO2} reports diagnostics computed on abnormal concentrations within the estimation window. Our goal was to identify the capacity of the models to adequately capture the spatio-temporal characteristics of NO$_2$ concentrations. Therefore, we focused on indicators of centrality, dispersion, shape and dependence of the AC$_{st}$ within the estimation window.
We considered the following set of diagnostics: Pearson's linear correlation index ($\rho$) for cross-sectional dependence; shape indices (i.e. skewness and kurtosis); sample mean ($\mu$); standard deviation ($\sigma$); temporal autocorrelation at various lags (i.e., $\phi_1$ is the autocorrelation coefficient 1-day lag, $\phi_3$ is the autocorrelation coefficient 3-day lag, $\phi_7$ is the autocorrelation coefficient at 7-day lag, $\phi_{14}$ is the autocorrelation coefficient at 14-day lag and $\phi_{21}$ is the autocorrelation coefficient at 21-day lag).
Finally, the Outliers quantity calculates the percentage of outliers over the total using the Hampel filter \citep{Hampel2016}. 
Statistics are calculated as the average value over the $N=83$ univariate time series.

\begin{table}[htbp]
\centering
\begin{tabular}{lcccc}
  \hline
 & HDGM & regARIMA & regAR1 & lm \\ 
  \hline
  $\rho$ & 0.02 & 0.37 & 0.34 & 0.35 \\ 
  $\mu$ & -0.00 & 0.02 & 0.01 & -0.00 \\
  $\sigma$ & 1.40 & 6.23 & 6.40 & 8.02 \\
  Skewness & 0.22 & 0.18 & 0.23 & 0.36 \\
  Kurtosis & 6.32 & 5.17 & 4.94 & 4.18 \\
  $\phi_1$ & -0.30 & -0.00 & -0.04 & 0.56 \\ 
  $\phi_3$ & -0.01 & 0.01 & 0.06 & 0.30 \\ 
  $\phi_7$ & 0.06 & 0.08 & 0.15 & 0.30 \\ 
  $\phi_{14}$ & 0.05 & 0.05 & 0.10 & 0.18 \\ 
  $\phi_{21}$ & 0.02 & 0.04 & 0.07 & 0.14 \\ 
  Outliers & 3.43 & 2.51 & 2.75 & 1.60 \\ 
   \hline
\end{tabular}
\caption{Descriptive statistics for abnormal concentrations ($AC$) of NO$_2$ within the estimation window (1st January 2018 - 8th March 2020). Each statistic is computed as the average value across the $N=83$ time series. $\rho$ is the Pearson's linear correlation, $\mu$ is the sample mean, $\sigma$ is the sample standard deviation, $\phi_l$ is the temporal autoccorelation coefficient at lag $l$ days,
and Outliers is the percentage of values classified as outliers.}
\label{tab:descript_AC_NO2}
\end{table}

The table suggests considerations that are consistent with Figure \ref{fig:ES_stats}. The HDGM residuals are (on average) cross-sectionally uncorrelated, while the other models show residual cross-sectional correlation. The very short-term (1-day) temporal serial correlation not zero for HDGM, while it is very close to zero for regARMA models. As the lag increases, the optimal regARMA model and HDGM keep the autocorrelation very low, while the regAR1 model and lm do not clean the residuals adequately. The variance of abnormal concentrations for HDGM is the lowest ever and is about 1/3 of the others, which confirms what the series graphs showed earlier (in which the range was the most compact of all). The shares of outliers are also very small.

In summary, we can state that the HDGM spatio-temporal model is extremely useful in solving the problem of spatial dependence by slightly sacrificing the very short term spatial autocorrelation (while the medium to long term spatial autocorrelation is preserved). This result may be due to the model definition itself. HDGM is built to model only the AR(1) dynamics. If the dependence is strong, covariates play a key role. In contrast, the regARIMA model with optimal choice of lags easily fits the high persistence by increasing the lags in the model. 

\subsection{Test statistics for NO$_2$ concentrations}
Table \ref{tab:ES_AQ_res} reports the estimated test statistics on NO$_2$ average concentrations during the event window. The reported significance levels are computed using the quantiles of a standard Normal distribution at level $\alpha$.

\begin{table}[htbp]
\centering
\begin{tabular}{llcccc}
  \hline
ES\_stats & CD & HDGM & regARMA & regAR1 & lm \\ 
  \hline
P1                  & Adj   & -16.69***     & -3.75***  & -9.39***  & -14.57*** \\ 
P2                  & Adj   & -20.57***     & -3.21***  & -8.60***  & -13.77*** \\ 
Corrado\_Tuckey     & Adj   & -22.76***     & -9.46***  & -15.12*** & -15.86*** \\ 
Z\_patell\_adj      & Adj   & -29.41***     & -11.32*** & -20.31*** & -22.95*** \\ 
Z\_BMP\_adj         & Adj   & -11.43***     & -2.24**   & -2.88***  & -2.83*** \\ 
T\_grank            & Adj   & -14.79***     & -11.9***  & -15.12*** & -15.26*** \\ 
Z\_grank\_adj       & Adj   & -26.37***     & -3.74***  & -9.9***   & -14.29*** \\ 
CumRank             & Adj   & -27.62***     & -3.92***  & -10.37*** & -14.97*** \\ 
CumRank\_mod        & Adj   & -79.69***     & -3.95***  & -11.07*** & -17.37*** \\ 
CumRank\_T          & Adj   & -112.38***    & -20.99*** & -56.94*** & -88.11*** \\ 
CumRank\_Z\_adj     & Adj   & -42.45***     & -3.68***  & -9.77***  & -16.08*** \\ 
cross\_T\_test      & Unadj & -25.81***     & -3.56***  & -9.43***  & -16.2*** \\ 
crude\_dep\_T\_test & Unadj & -13.75***     & -15.07*** & -25.86*** & -23.39*** \\ 
T\_test\_skew       & Unadj & -210.02***    & -20.51*** & -54.54*** & -101.06*** \\ 
Z\_patell           & Unadj & -60.82***     & -3.66***  & -10.07*** & -18.41*** \\ 
CumRank\_Z          & Unadj & -39.70***     & -3.96***  & -10.81*** & -16.24*** \\ 
Z\_grank            & Unadj & -5.22***      & -2.24**   & -2.87***  & -2.82*** \\ 
Z\_BMP              & Unadj & -7.94***      & -1.61 .   & -3.04***  & -3.37*** \\ 
\hline
\end{tabular}
\caption{Estimated ES statistics for NO$_2$ concentrations within the event window (9th March 2020 - 18th May 2020). Symbols '***', '**', '*' stand for statistically significant at levels 1\%, 5\% and 10\%, respectively. The symbol '.' stands for not statistically significant at 10\%.}
\label{tab:ES_AQ_res}
\end{table}

Overall, the estimated values are consistent with the expectations, both in terms of sign and magnitude.
Regarding the signs, all the estimated statistics report a negative value, meaning the presence of a reduction in the average NO$_2$ concentrations at regional level due to the lockdown period. Estimations for HDGM, regAR1 and lm are always statistically significant at 1\% level. The statistical significance for the optimal regARMA are slightly lower, especially for the unadjusted statistics (e.g., $Z_{BMP}$ and $Z_{grank}$). This result is consistent with the findings observed in Figure \ref{fig:ES_stats}, which showed that the difference between the averages of the two periods was very close to zero.
Eventually, considering the HDGM, in absolute values, all the unadjusted statistics are smaller than the cross-sectionally adjusted statistics. This finding is also consistent with expectations. Having explicitly modeled cross-sectional dependence in estimation, which led to a near-zero Pearson correlation index in the estimation window, the adjusted statistics result in very high values by construction. However, adjusting the ES statistics remains the most consistent and statistically valuable alternative.

\section{Conclusions and future developments} \label{sec:ES_conclusions}
We presented new empirical findings regarding the application of ES methods to the case of strongly cross-sectional dependent multivariate time series. Our interest focused on cross-sectional dependence resulting from the strong spatial correlation naturally exhibited by geo-referenced observations, such as air quality measurements collected from ground monitoring networks. 

The major points we raised were as follows.
First, we generalized the statistical concepts underlying ES and adapted them to the case of airborne pollutant concentrations. In this regard, we provided definitions of normal and abnormal airborne concentrations by exploiting the analogy with the concept of abnormal returns typical of financial literature.
Second, we discussed the role of (spatial) cross-sectional dependence in ES by proposing a twofold strategy to overcome its drawbacks. In principle, we proposed to analyze time series interconnected by a strong spatial correlation through a geostatistical spatio-temporal model, namely the HDGM. This model allows to exploit the relationship among the response variable and a set of exogenous covariates, while accounting for the spatio-temporal dynamics of the observations. Additionally, the HDGM was compared with three purely temporal models not explicitly accounting for the spatial dimension. As a second step, we suggested to investigate the abnormal concentrations through a set of sixteen ES statistics in order to determine the significance of the event of interest. Among the ES statistics used, some of them were specifically built to deal with cross-sectionally dependent observations.

The adjustment strategy was applied to the empirical case of NO$_2$ concentrations registered in Lombardy through the ARPA air quality monitoring network. We considered as the event of interest the lockdown restrictions imposed on mobility and socio-economic activities during the first wave of the COVID-19 pandemic. The main interest was to state if the lockdown generated significant reductions in the average concentrations of NO$_2$, i.e., we tested for a level shift after the event date.
Our findings can be summarized as follows. First, the HDGM was the only one that distinctly reveals the level shift caused by the event of interest. The other models, in fact, showed much more limited reductions. Moreover, the abnormal concentrations of the HDGM were the least volatile among those computed. Second, the HDGM was the only one of the four specifications to fully isolate the spatial dependence of the data, leading the average CD of the abnormal concentrations to be zero. However, the very short-term temporal autocorrelation (i.e., 1-day lag) persisted. Third, all the proposed test statistics, both adjusted and unadjusted for CD, unequivocally revealed a reduction in NO$_2$ concentrations due to the lockdown imposed in Lombardy. The robustness of this reduction was established through a sensitivity analysis of the test statistics with respect to the start and end dates of the event.

Future studies that apply Event Studies methodology for spatio-temporal data should regard the spatio-temporal dependence between observations intensively.
We focused on modelling NO$_2$ concentrations using the univariate HDGM. However, other models could be implemented. Consider, for example, the fixed rank smoothing spatio-temporal random effects model \citep{KatzfussCressie2011}, hierarchical spatio-temporal models based on INLA \citep{FioravantiEtAl2021,SaezBarcelo2022} and Dynamic Spatio-Termporal Models \citep{ZammitMangion2019}.
The second strategy involves the use of multivariate models to jointly analyze multiple response variables (e.g., different types of pollutants) that are mutually correlated. This approach would provide an opportunity to exploit the cross-correlation among responses to further improve predictions in the event window. Consider as examples the multivariate extension of HDGM \citep{SPASTA2021} or spatio-temporal VAR models \citep{FerreiraMateuPorcu2022}.
Eventually, the ES test statistics could be explicitly adjusted for the spatial cross-correlation \citep{Chen2015} and spatio-temporal cross-correlation \citep{MaEtAl2006,GaoEtAl2019} measures, instead of linear correlation measures.


\section*{Data and codes}
All the results presented in this paper can be reproduced using R and Matlab software. Data and codes have been uploaded in the journal's database. Data and codes are available at the following Google Drive link: \href{https://drive.google.com/drive/folders/1fXImuCvOpP7orvfuSeGwDa-GcuNyOf-H?usp=sharing}{https://drive.google.com/drive/folders/1fXImuCvOpP7orvfuSeGwDa-GcuNyOf-H?usp=sharing}.

\section*{Supplementary materials}
This paper is accompanied by a single supplementary material: Full results of the event date sensitivity analysis (S1).

\section*{Declarations}
\begin{itemize}
\item Funding: this research was funded by Fondazione Cariplo under the grant 2020-4066 "AgrImOnIA: the impact of agriculture on air quality and the COVID-19 pandemic" from the "Data Science for science and society" program.
\item Conflict of interest/Competing interests: the authors declare that they have no competing interests
\end{itemize}

\newpage

\end{document}